\documentclass[twocolumn,showpacs,preprintnumbers,amsmath,amssymb]{revtex4}

\usepackage{graphicx}
\usepackage{dcolumn}
\usepackage{bm}
\usepackage{epsfig}

\newcommand{\mum}{$\mathrm{\,\mu m}$}

\newcommand{\muK}{$\mathrm{\,\mu K}$}

\newcommand{\BE}{Bose-Einstein}
\newcommand{\BEC}{Bose-Einstein condensate}
\newcommand{\BECs}{Bose-Einstein condensates}
\newcommand{\IP}{Ioffe-Pritchard}

\begin{document}


\title{Bose-Einstein condensation in a magnetic double-well potential}

\author{T.G. Tiecke}
\author{M. Kemmann}
\author{Ch. Buggle}
\author{I. Shvarchuck}
\author{W. von Klitzing}
\author{J.T.M. Walraven}


\affiliation{FOM Institute for Atomic and Molecular Physics\mbox{,} Kruislaan 407, 1098 SJ Amsterdam, The Netherlands}

\date{\today}

\begin{abstract}
We present the first experimental realisation of Bose-Einstein condensation in a purely magnetic double-well potential.  This has been
realised by combining a static Ioffe-Pritchard trap with a time orbiting potential (TOP).  The double trap can be rapidly switched to a
single harmonic trap of identical oscillation frequencies thus accelerating the two condensates towards each other.  Furthermore, we show
that time averaged potentials can be used as a means to control the radial confinement of the atoms.  Manipulation of the radial
confinement allows vortices and radial quadrupole oscillations to be excited.
\end{abstract}



\maketitle

\section{\label{sec:Intro}Introduction}

Bose-Einstein condensates in dilute vapours (BEC) have been studied using magnetic
\cite{AndersonCornellSci95,BradleyHuletPRL95,DavisKetterle95} and optical \cite{BarrettChapmanPRL01,StamperKurnKetterlePRL98} potentials.
Magnetic traps can be divided into two classes: static and dynamic traps.  \BE\ condensation has first been observed in a dynamic trap
using a time orbiting potential (TOP) \cite{AndersonCornellSci95} and shortly thereafter in a static \IP\ trap \cite{DavisKetterle95}.  A
TOP trap uses a three dimensional quadrupole field to confine the atoms.  In order to avoid Majorana losses near the centre of the trap a
rotating homogeneous field shifts the field minimum onto a circle around the trapped sample.  At an appropriate rotation frequency the
trapping potential is simply the time average of the orbiting potential.  The \IP\ trap on the other hand is designed to have a static,
non-zero field minimum: in two directions ({\it x,y}) the atoms are confined by a quadrupole field.  On the {\it z}-axis the radial field
component is zero and the axial component is a parabola $B_z = B_{0} + \beta z^2$ with a curvature $\beta$ and an offset $B_{0}$.

\begin{figure}[fb]
  \begin{center}
  \psfig{figure=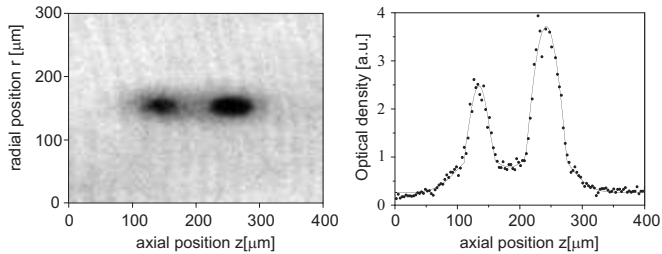,width=8.7cm}
  \end{center}
  \caption{Absorption image of two \BECs\ created in a double TOP trap.  The condensates have been accelerated towards each other by
  switching to a single trap.  The picture is taken after 3\,ms of free expansion.  The figure on the right is an optical density profile through the centre of both condensates.  The solid line is a
  fit to the experimental data.}
  \label{fig:DoubleCloud}
\end{figure}

In this paper we explore the possibilities offered by time averaged potentials (TAP) in \IP\ traps.  We distinguish two cases $B_{0}>0$ and
$B_{0}<0$.  In the first case $B_z$ is always positive and the axial confining potential is a parabola.  We can influence the radial and
axial confinement of the atoms by applying a {\it linear} TAP-field, i.e.\ a homogeneous modulation field orthogonal to the {\it z}-axis.
This allows us to introduce a radial ellipticity in the otherwise axially symmetric trap, which can be exploited to drive vortices or
quadrupole oscillations.  In the second case ($B_0<0$),  $B_z$ crosses zero at two points on the {\it z}-axis creating an axial double well
potential.  We can create a double TOP trap by adding a {\it circular} TAP-field orthogonally to the {\it z}-axis
\cite{AndersonCornellSci95,ThomasFoot02}.  As an example we demonstrate \BE\ condensation in a double TOP trap (see figure
\ref{fig:DoubleCloud}).

\section{\label{sec:Method}Trap geometry}

We explored the possibilities offered by TAP fields in a \IP\ coil configuration using an efficient semi-analytic model of the trap.  We
calculate the magnetic field for the detailed geometry of each coil separately over a three-dimensional grid around the trap centre.  Two
different regions are chosen: one for the cold atom cloud and the \BEC\ ($10 \times 10 \times 100$\mum) and one for the hot thermal cloud
($1 \times 1 \times 10$\,mm).  On this grid a three-dimensional fourth-order polynomial is fitted for each of the coils.  The final field
is calculated as the sum over the polynomials with the electrical currents of the individual coils as parameters.  The frequencies and
nonlinearities of the trap are then simply the appropriate coefficients of the final polynomial.  Using these polynomials \emph{any} field
possible with our coil configuration can be rapidly calculated.  For selected cases we also derive simplified analytical expressions to
emphasise the principles involved.

In order to elucidate the TAP principle we write the potential $U(x,y,z)$ of the axially symmetric ($\omega_\rho\equiv\omega_x=\omega_y$)
\IP\ trap in the approximation of a elongated trapping potential $(\omega_\rho \gg\,\omega_{z})$:
\begin{eqnarray}
  U(x,y,z)&=&\mu\,\sqrt{\alpha^2(x^2+y^2) + \left(B_{0}+\frac{1}{2}\beta z^2\right)^2}
  \label{equ:SingleTrapPotential}
\end{eqnarray}
where $\alpha$ is the radial gradient, $\mu$ the magnitude of the magnetic moment of the atoms and $m$ the atomic mass.  For the standard
\IP\ trap $(B_{0}>0)$ the harmonic frequencies are
\begin{eqnarray}
  \omega_{\rho} =\sqrt{\frac{\mu}{m}\,\frac{\alpha^2}{B_{0}}}~,~~~
  \omega_{z}=\sqrt{\frac{\mu}{m}\,\beta}\ \label{equ:SingleTrapFrequency} \ .
\end{eqnarray}

\subsection{Magnetic double-well potential ($B_0<0$)}

For a negative offset the modulus of the field becomes zero at two points on the {\it z}-axis creating two three-dimensional quadrupole
traps.  The distance ${\tiny \Delta} z$ between the two trap centres can be controlled via $B_0$,
\begin{equation}
 {\tiny \Delta z} = 2 \sqrt{2 |B_{0}|/\beta }\ . \label{equ:Splitting}
\end{equation}
Inversely, we can conveniently determine $B_0$ from a measurement of $\tiny \Delta z$.

To eliminate depolarisation near the trap minima we use a circular TAP, creating a double TOP configuration
\cite{ThomasFoot02,PetrichCornellPRL95}.  A homogeneous field of amplitude $B_m$ orthogonal to the {\it z}-axis displaces the magnetic
field zero by a distance
\begin{equation}
\rho_m=B_m/\alpha~\label{equ:RadialDisplacement}.
\end{equation}
This expression is conveniently used to calibrate $B_m$.

Rotating the modulation field around the {\it z}-axis (with an angular modulation frequency $\omega_m$) moves the zero along a so-called
`circle of death'.  If the modulation frequency is large compared to the oscillation frequency of the atoms in the trap but slow compared
to the Larmor frequency (\,$\omega_{\rho,z} \ll \omega_m \ll \mu B_m / \hbar$\,) the atoms will see the time-average of the modulated
potential \cite{PetrichCornellPRL95}:
\begin{equation}
  U_{\text{TAP}}(x,y,z)=\frac{\omega_m}{2 \pi}\int_{0}^{\frac{2\pi}{\omega_m}}U(\rho_m\sin \omega _{m}t,\,\rho_m\cos \omega _{m}t,\,z)dt
  .\nonumber
\end{equation}
The harmonic trapping frequencies for both minima of the double-well potential are
\begin{equation}
\overline{\omega}_\rho= \sqrt{\frac{\mu}{m}\frac{\alpha^2}{2B_m}}~,~~~
\overline{\omega}_z=\sqrt{\frac{\mu}{m}\frac{2\beta\,|B_{0}|}{B_m}}\label{equ:SplitTrappingFrequencies}~.
\end{equation}
Note that $\omega_z=0$ for $B_{0}=0$ and $B_m > 0$.  For this case the axial confinement is governed purely by higher order terms.

The two traps are separated by a potential barrier of
\begin{equation}
U_{\text{barrier}}=\mu\left(\sqrt{B_{0}^2+B_m^2}-B_m\right)
\end{equation}
where the first term in the brackets corresponds to the field at the origin and $B_m$ to the field minimum of each of the wells.

One of the interesting aspects of this double trap is the possibility of rapidly switching from the double TOP trap to a static \IP\ trap.
This is done by removing the TAP field and switching from $B_0 < 0$ to $B_0' > 0$, changing neither $\alpha$ nor $\beta$.  The condensates
then accelerate towards each other and collide in the origin at a {\it relative} kinetic energy of $E_{\text{kin}}=4 \mu |B_{0}|$.  In a
harmonic trap the shape of a cloud is completely decoupled from its centre-of-mass motion.  Therefore, if the trapping frequencies in the
single and double traps are equal, the gas clouds move towards each other without changing shape.  It follows from equations
\ref{equ:SingleTrapFrequency} and \ref{equ:SplitTrappingFrequencies} that the axial \emph{and} radial frequencies of the double and single
traps are the same if the offset and modulation fields obey
\begin{equation}
B_0'=2 B_m= -4 B_0~.
\end{equation}

In our trap we can easily explore collision energies up to 1.3\,mK corresponding to 7000 recoil energies at 780\,nm.

Let us turn to the question of gravity.  If the trap is slightly tilted gravity will affect the relative depth of the two potential minima
and thus the relative number of atoms in the two traps.  As the distance between the traps increases the difference in atom numbers
continues to grow.  If the traps were to remain coupled all atoms would eventually collect in the lower trap.  Therefore the two traps have
to separated sufficiently fast.

\subsection{Radial Ellipticity ($B_0>0$)}

For a positive offset we have a standard \IP\ potential.  If we then apply a linear TAP in the {\it x}-direction modulating at an angular
frequency $\omega_m$ we can write the time-averaged potential as
\begin{equation}
U_{\text{TAP}}(x,y,z)=\frac{\omega_m}{2 \pi}\int_{0}^{\frac{2\pi}{\omega_m}}U(\rho_m~\sin \omega_m t,\,y,\,z)dt~.
\end{equation}
The radial harmonic trap frequencies ($\overline{\omega}_x$, $\overline{\omega}_y$) can be expressed in terms of the ratio $b \equiv
B_{m}/B_{0}$:
\begin{eqnarray}
\overline{\omega}_x =\omega_{\rho} \sqrt{\frac{1}{1+b^2}\mathsf{E'}(b)}~,~~~ \overline{\omega}_y =\omega_{\rho}
\sqrt{\mathsf{K'}(b)}\label{equ:Ellipticity}~,
\end{eqnarray}
where  $\mathsf{K'}(b)=\frac{1}{2 \pi}\int_{0}^{2 \pi}(1+b^2 \sin^2 \hspace{-1mm} \phi)^{-1/2} \,d\phi$ and $\mathsf{E'}(b)=\frac{1}{2
\pi}\int_{0}^{2 \pi}(1+b^2 \sin^2 \hspace{-1mm} \phi)^{1/2}\,d\phi$ \cite{EllipticalIntegral}.  For $b\ll 1$ the ellipticity is given by
\begin{equation}
\epsilon\equiv\frac{\overline{\omega}_x^2-\overline{\omega}_y ^2}{\overline{\omega}_x ^2+\overline{\omega}_y ^2}\approx -b^2/4.
\end{equation}
For $b=1$ we have $\epsilon=-0.16$\,.

Radial ellipticity can be used for example to excite quadrupole oscillations or to create vortices in a \BEC.  To drive the radial
quadrupole oscillation the elipticity of the trap has to be modulated for example by driving the fields as $B_x(t)=B_m \sin \omega_m t
\,(1+\sin \Omega t)/2$ and $B_y(t)=B_m \cos \omega_m t \,(1-\sin \Omega t )/2$, where $\Omega$ is matched to the frequency of the
quadrupole oscillation.

Vortices can be created by rotating an elliptic trapping potential at an angular frequency $\Omega$.  This has been demonstrated using
rotating laser beams in a magnetic trap \cite{MadisonDalibard00,AboShaeerKetterleSCI01} and in an asymmetric TOP trap
\cite{HodbyFoot02,HaljanCornell01}.  We can do this by rotating the linear TAP-field as $B_x(t)=B_m \sin\omega_m t \,\sin \Omega t$ and
$B_y(t)=B_m\sin \omega_m t \,\cos \Omega t$.

\section{BEC in a double TOP}

As an example of the use of time-averaged potentials in combination with a \IP\ coil configuration we demonstrate \BE\ condensation in a
double TOP trap.

\begin{figure}[fb]
  \begin{center}
  \psfig{figure=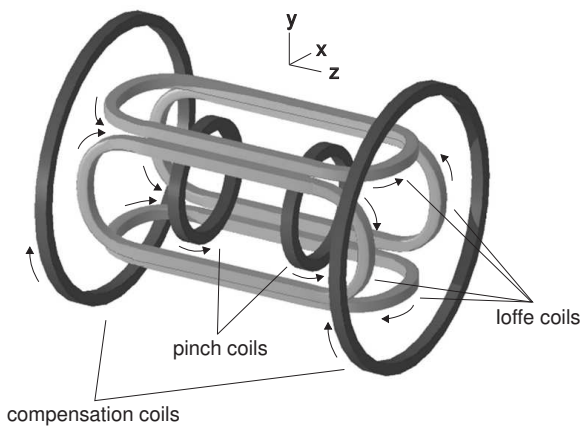,width=8.5cm}
  \end{center}
  \caption{The coil configuration of our Ioffe-Pritchard trap.  The arrows indicate the direction of the currents in
  the main coils.  Modulation coils (not shown) are attached to the outside faces of the Ioffe coils.}
  \label{fig:TheTrap}
\end{figure}

Our \IP\ trap is shown in figure \ref{fig:TheTrap} and is described in detail in reference \cite{Dieckmann01}.  The trap consists of four
Ioffe coils producing the radial quadrupole field.  The field of the pinch coils provides the axial parabolic potential.  The compensation
coils are designed to compensate the field of the pinch coils at the origin.  Both sets of coils are connected in series in order to reduce
field noise.  The offset $B_0$ is controlled by applying a dedicated homogeneous field (coils not shown in figure \ref{fig:TheTrap}).  In
order to be able to create the modulation fields we have attached modulation coils to the outer faces of the Ioffe coils.  They consist of
PCB boards with one layer of 35\mum\ copper on either side, connected in series.  In order to maximise the trap stability facing modulation
coils were driven in series.

The experimental path towards BEC into the magnetic double-well is as follows.  We optically pump about $6\times 10^9$ atoms from a
$\mathrm{^{87}Rb}$ MOT into the $|5^{2}S_{1/2},F=2,\ m_{F}=2\rangle $ state.  These are loaded into a matching, horizontal \IP\ trap
($\omega _{\rho }=2\pi \times 8 $\,Hz, $\omega _{z}=2\pi \times 7$\,Hz and $B_{0}=37\,$G).  The circular TAP-field of $B_m=0.68(3)$\,G
 rotating at a frequency of $7.0$\,kHz  remains constant throughout the experiment \cite{AudioAmp}.  The transfer efficiency is about 50\% and
the temperature in the trap is about 70\muK.  We then compress the trap to $\omega _{\rho }=2\pi\times 390(10)$\,Hz and $\omega _{z}=2\pi
\times 14.8(4)$\,Hz by ramping the currents in the main coils from 50\,A up to their final value of 400\,A and reducing the offset to
$B_{0}=0.4$\,G.  At this current we have $\alpha=352(1)$\,G\,cm$^{-1}$ and $\beta=266(2)$\,G\,cm$^{-2}$ \cite{ShvarchuckWalravenXXX02}.

Rapid forced evaporation using a radio-frequency sweep-down to an intermediate value of $433$\,kHz above the bottom of the trap cools the
sample to a temperature of about $8$\muK.  At this temperature 90\% of the atoms lie within the future circle of death.  We then ramp the
offset-field $B_{0}$ linearly down at a rate of $-0.5\,\mathrm{G\,s^{-1}}$ eventually splitting the cloud.

\begin{figure}[fb]
\vspace{0.2cm}
  \begin{center}
  \psfig{figure=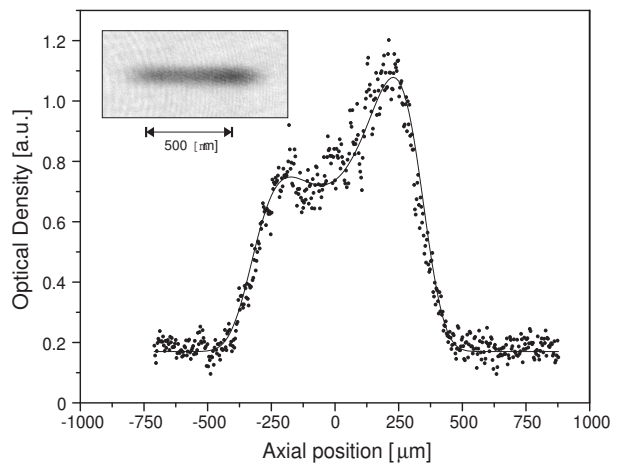,width=8cm}
  \end{center}
  \caption{An optical density profile along the {\it z}-axis of the thermal cloud during the early stages
  of separation after a free expansion of 4\,ms.  The cloud contains  $10^6$ atoms at a temperature of
  0.66\muK\ as determined from the radial expansion.  The solid line is a prediction from the
  model with the angle  ($0.4$ degrees) and maximal optical density as adjustable parameters.  The inset
  shows the absorption image.}
  \label{fig:ThermalSplit}
\end{figure}

Figure \ref{fig:ThermalSplit} shows a profile through a thermal cloud of $10^6$ atoms at a temperature of 0.66\muK\ during the early stages
of the splitting process at $B_0=-53(1)$\,mG.  The temperature and atom number was determined by time-of-flight imaging.  The solid curve
is calculated for the measured temperature using the semi-analytical model adjusting the maximum optical density and the tilt angle (0.4
degrees) to fit the data.

\begin{figure}[ft]
  \begin{center}
  \psfig{figure=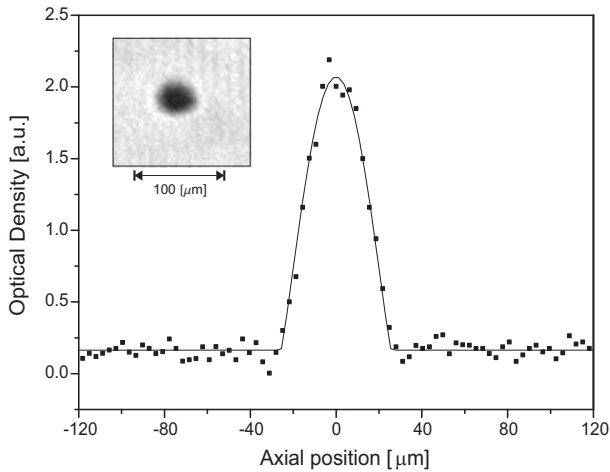,width=8cm}
  \end{center}
  \caption{An optical density profile along the {\it z}-axis of one of the two \BECs\ after 4\,ms of
  free expansion.  The solid line is a fit
  of a Thomas-Fermi profile to the experimental points.  The \BEC\ contains $4\times 10^4$ atoms.  The inset
  shows the absorption image.}
  \label{fig:BECcut}
\end{figure}

After the cloud is fully split we fix the offset at $B_0=-0.63(1)$\,G and condense the sample by lowering the radio-frequency to 23\,kHz
above the bottom of the trap.  Figure \ref{fig:BECcut} shows a profile of an absorption image of one of the two \BECs\ after a free
expansion of 4\,ms.  The solid line is a fit to the Thomas-Fermi distribution in a harmonic trap.  The condensate contains $4\times 10^4$
atoms.

\begin{figure}[fh]
  \begin{center}
  \psfig{figure=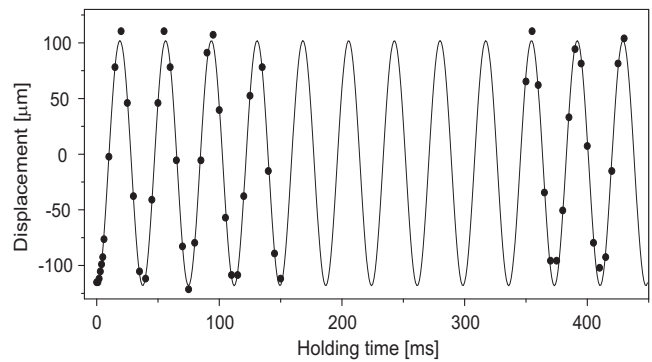,width=8.5cm}
  \end{center}
  \caption{An axial centre-of-mass oscillation of a \BEC\ in a double-well trap.  The solid line
  corresponds to a fit of a sine function to the data ($\omega_z=2\pi\times26.8$\,Hz).}
   \label{fig:MarbleOscillation}
\end{figure}

 We excited a centre-of-mass oscillation of a \BEC\ along the {\it z}-axis by jumping the offset $B_{0}$ from
-0.43(1)\,G to -0.63(1)\,G thus shifting the traps by 115(1)\mum\ outwards.  The condensates started to oscillate around the new trap
centre (dots in figure \ref{fig:MarbleOscillation}).  The solid line is a fit of a sine wave to the data ($\omega_z=2\pi\times26.8$\,Hz).
From our semi-analytical model including an anharmonic shift of 0.8\,Hz we find a frequency of 27.5(8)\,Hz.  Hence we have agreement with
the data within the experimental error.

\section{conclusions and outlook}

We explored the TAP principle as a powerful tool to manipulate \IP\ traps.  For a positive $B_0$ a radial ellipticity can be induced and
used to excite vortices and quadrupole oscillations.  For a negative $B_0$ double well potentials are obtained.  These can readily be
converted to a single trap and thus used in collision experiments with ultra-cold gas clouds.  As an example we used a circular TAP field
to demonstrate \BE\ condensation in a double TOP trap.

This work is part of the Cold Atoms program of FOM.

\bibliography{DoubleTrap}

\begin{thebibliography}{10}
\bibitem{EllipticalIntegral}
Note that the integrals $\mathsf{E'}$ and $\mathsf{K'}$ used in equation
  \cite{equ:Ellipticity} are related to the elliptical integrals $\mathsf{E}$
  and $\mathsf{K}$ as $\mathsf{E'}(b)= 2 \mathsf{E}\left(-b^2\right)/ \pi$ and
  $\mathsf{K'}(b)= 2 \mathsf{K}\left(-b^2\right)/ \pi$.
\bibitem{AudioAmp}
The modulation current of ${I}\approx3$\,{A} corresponing to the modulation
  field of ${B}_m=0.68(3)$\,{G} was driven by a standard audio amplifier
  ({Y}amaha {AX}-496).
\end{thebibliography}
\bibliographystyle{unsrt}

\end{document}